\documentclass[a4paper,fleqn]{cas-sc}

\usepackage[ruled,vlined]{algorithm2e}
\usepackage{enumitem}
\usepackage{cite}
\usepackage{graphicx}
\usepackage{subfig}
\usepackage{multirow}
\usepackage{subcaption}
\usepackage{diagbox}
\usepackage[authoryear,longnamesfirst]{natbib}

\def\tsc#1{\csdef{#1}{\textsc{\lowercase{#1}}\xspace}}
\tsc{WGM}
\tsc{QE}
\tsc{EP}
\tsc{PMS}
\tsc{BEC}
\tsc{DE}

\begin{document}
\let\WriteBookmarks\relax
\def\floatpagepagefraction{1}
\def\textpagefraction{.001}

\shorttitle{Decentralized Federated Anomaly Detection in Smart Grids: A P2P Gossip Approach}

\shortauthors{M. A. Husnoo et~al.}

\title [mode = title]{Decentralized Federated Anomaly Detection in Smart Grids: A P2P Gossip Approach}                      
\tnotemark[1]

\tnotetext[1]{This document is one of the results of the research project funded by the Centre for Cyber Resilience and Trust (CREST), School of Information Technology, Deakin University.}

%

\author[1]{Muhammad Akbar Husnoo}[orcid=0000-0001-7908-8807]
\cormark[1]
\ead{mahusnoo@deakin.edu.au}

\affiliation[1]{organization={Centre for Cyber Resilience and Trust (CREST), Deakin University},
    addressline={75 Pigdons Rd}, 
    city={Waurn Ponds},
    postcode={3216}, 
    state={Victoria},
    country={Australia}}

\author[1]{Adnan Anwar}[orcid=0000-0003-3916-1381]
\ead{adnan.anwar@deakin.edu.au}

\author[2]{Md Enamul Haque}[orcid=0000-0002-8893-2181]
\ead{enamul.haque@deakin.edu.au}

\affiliation[2]{organization={Centre for Smart Power and Energy Research (CSPER)), Deakin University},
    addressline={75 Pigdons Rd}, 
    city={Waurn Ponds},
    postcode={3216}, 
    state={Victoria},
    country={Australia}}

\author[3]{Abdun Naser Mahmood}[orcid=0000-0001-7769-3384]
\ead{a.mahmood@latrobe.edu.au}

\affiliation[3]{organization={Department of Computer Science \& IT, Latrobe University},
    addressline={Plenty Rd}, 
    city={Bundoora},
    postcode={3086}, 
    state={Victoria},
    country={Australia}}

\cortext[cor1]{Corresponding author}

\begin{abstract}
Amidst escalating concerns regarding security and privacy within the Smart Grid domain, the need for robust intrusion detection mechanisms in critical energy infrastructure has surged in recent times. To address the challenges posed by privacy preservation and decentralized power zones with distinct data ownership, Federated Learning (FL) has emerged as a promising privacy-preserving solution which facilitates collaborative training of attack detection models without necessitating the sharing of raw data. However, FL presents several implementation limitations in the power system domain due to its heavy reliance on a centralized aggregator and the risks of privacy leakage during model update transmission. In response to the technical bottlenecks, this paper introduces a novel decentralized federated anomaly detection scheme based on two main gossip protocols namely Random Walk and Epidemic. Our findings indicate that the Random Walk protocol exhibits superior performance compared to the Epidemic protocol, highlighting its efficacy in decentralized federated learning environments. Experimental validation of the proposed framework utilizing publicly available industrial control systems datasets demonstrates superior attack detection accuracy while safeguarding data confidentiality and mitigating the impact of communication latency and stragglers. Moreover, a notable 35\% improvement in training time against conventional FL highlights the efficacy and robustness of our decentralized learning approach.
\end{abstract}




\maketitle

\section{Introduction}

The large-scale prevalence of Internet of Things (IoT) technology and cutting-edge communication protocols within the Smart Grid (SG) paradigm has provoked a significant surge in data volumes emanating from edge devices. Such vital power-related data play a crucial role in various key operations of the grid including demand management, grid monitoring and control, fault detection and isolation, etc. Consequently, the management of the overall grid infrastructure can be carried out more dynamically and efficiently. However, the increasing heterogeneity, diversity, and complexity within the power grid ecosystem present significant challenges to its integrity. In modern times, ensuring data integrity is a priority for modern power systems which is due to the ubiquity of modern cyber-physical power grid components alongside advancements in information technology, which now enable data processing and decision-making at the edge, in contrast to traditional cloud storage-based control centers. 

On another hand, whilst the digitisation of the modern-day power systems has managed to alleviate the response issues to the insatiable demand for energy, smart grids are still prone to several technical bottlenecks. Empowered by the massive amounts of sensitive data generated, a wide range of the cyber-physical components of the SG has become potential zero-day vulnerability susceptible to cyber threats by malicious actors, including but not limited to False Data Injection Attacks (FDIA), Time Synchronization Attacks, etc. Hence, it is preemptive to leverage state-of-the-art technological advances to develop effective countermeasures to defend SGs against such types of intrusions. Consequently, the use of Machine Learning and Deep Learning (DL) emerged as driving forces behind several breakthroughs in grid sabotage detection due to their prowess at identifying relevant complex patterns from diverse data sources. 

Despite their remarkable contributions, such focalized anomaly detection schemes suffer from several shortcomings which include limited storage capabilities, communication network overhead and, foremost, privacy issues. Experts frequently disregard the fact that grid information encompasses sensitive data which carries potential privacy risks during transfer to centralized repositories for analytical tasks. Thus, to tackle such impediments, training deep learning models in a distributed fashion, essentially Federated Learning (FL) garnered momentum as a plausible privacy-preserving alternative which enables edge-based on-device training bereft of sharing raw data thus mitigating the associated privacy risks. For instance, one of our previous works \cite{feddisc} in similar line proposed a computationally-efficient privacy-preserving federated scheme to differentiate between anomalous events and system disturbances. Other studies \cite{9478223, 9531953, huong2021detecting} have similarly leveraged FL techniques for misbehaviour detection within SG ecosystems. Such literature have explored the effectiveness of FL at detecting irregularities and deviations from normal operating conditions, thus, reinforcing the security and resilience of power grids. 

\begin{figure}
    \centering
    \includegraphics[width=11cm]{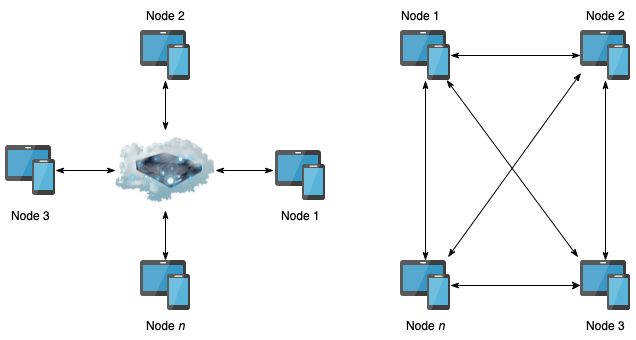}
    \caption{An illustrative comparison between conventional Federated Learning (FL) and Decentralized Federated Learning (DFL) provides insight into their fundamental differences and operational paradigms.}
    \label{fig:illusvsvs}
\end{figure}

However, traditional FL-based intrusion detection schemes, whilst being focussed on its client-server-centric design principles as shown in Fig. \ref{fig:illusvsvs}A, exhibits certain drawbacks that impede its practical implementation within real-world power system scenarios. One such limitation is the blind reliance of client nodes on a centralized aggregator to coordinate the learning process and aggregate local model updates. This may eventually lead to potential privacy risks \cite{10173657} and communication bottlenecks \cite{feddisc}, especially in large-scale and geographically distributed environments. Furthermore, traditional  FL approaches also lack behind in scalability, due to the centralized server becoming a single point of failure and expensive computational overhead with increasing number of client nodes. Such weaknesses motivate the exploration of Decentralized Federated Learning (DFL) whereby model aggregation and coordination occur both locally at the client nodes rather than requiring a central server, hence, reducing privacy concerns, improving scalability, and enhancing the robustness of distributed learning-based anomaly detection scenarios.

In this work, we advocate for the effectiveness of a decentralized federated approach to misbehaviour detection in modern power systems in view of discriminating between malicious and natural power system irregularities, thereby providing enabling reliable operational support to energy grid operators. Specifically, the main contributions of this manuscript are laid out in the following: 1) We leverage the capabilities of state-of-the-art hybrid anomaly detection neural networks which combines the strength from both sequential modelling and representation learning, specifically a Transformer-Autoencoder (TAE) model to offer accurate and timely detection of intentionally crafted malicious events from natural power system disturbances. 2) To the best of our knowledge, we propose the first gossip-based decentralized federated misbehavior detection approach which harnesses the collective intelligence of grid regions within a geographically decentralized power grid system while eliminating the need for a centralized aggregator, thereby mitigating the associated privacy concerns and scalability pitfalls associated with conventional federated anomaly detection methods. 3) To further optimize the computational overhead of our proposed solution, we employ a differentially-private gradient quantizer, particularly DP-SIGNSGD, which significantly reduces the byte-size of the model updates propagated through the network whist improving its privacy aspects. 4) Lastly, we perform comprehensive empirical evaluations of our proposed frameworks on openly accessible  Industrial Control Systems dataset to validate its performance computational overhead which yields a 94.2\% accuracy and 35\% cutback in training time against a conventional FL setup.

The rest of this manuscript is structured as follows: In Section \ref{sec:relatedliterature}, we briefly review the existing literature in relation to conventional FL and DFL. The proposed methodology, detection module and system model are thoroughly discussed in Section \ref{sec:proposedmethod}. Simulation scenarios to evaluate the effectiveness of our proposed DFL scheme on publicly obtained databases are presented in Section \ref{sec:results}. Finally, we conclude our article in Section \ref{sec:conclusion}.


\section{Related Literature}
\label{sec:relatedliterature}

In what follows, we discuss the current state-of-the-art on anomaly detection in power systems by grouping existing literature into two main categories as in the following:

\subsection{Conventional FL-based Anomaly Detection \& its Limitations}

Since its inception, FL has positioned itself as an emerging discipline that has achieved promising results in various SG-related applications whereby privacy is paramount. Taïk and Cherkaoui \cite{9148937} introduced FL to the SG domain by training a LSTM model on a  real-world Texas load consumption data to highlight the effectiveness of FL and its privacy-preserving guarantees in a demand load forecasting scenario. Another work in \cite{10072290} proposed a horizontal federated load forecasting solution for Retail Energy Providers and achieved sufficient forecasting performance using the Solar Home Electricity Data from Ausgrid. Following its success, multiple studies including \cite{ Fekri_Grolinger_Mir_2022, 9888131, Gholizadeh_Musilek_2022} continued in similar directions. 

Given its prominence and optimal results within demand forecasting, FL swiftly gained momentum in other facets of power systems, in particular, anomaly detection. For instance, researchers in \cite{9531953} developed a federated temporal convolutional network for identifying power theft. Through comprehensive data-driven experiments using a real-energy consumption dataset, they asserted that their federated framework can attain high detection accuracy with reduced computational overhead. Similarly, Ashraf et al. \cite{Ashraf_Waqas_Abbas_Baker_Abbas_Alasmary_2022} attained comparable performance for energy theft detection by validating federated conventional machine learning algorithms on real-world datasets. The works in \cite{feddisc, artman_akbar} proposed computationally-efficient differentially-private federated schemes to differentiate between anomalous events and system irregularities. Furthermore, authors in \cite{9547719, lin2022incentive, 9878267} developed FL-based frameworks for the detection of False Data Injection attacks in power systems with sufficient performance. More recently, to mitigate the effects of straggler client nodes and communication inefficiency in relation to conventional FL, the works in \cite{10226030, 10.1145/3579856.3592824, 10437420} proposed semi-asynchronous FL-based anomaly detection frameworks to discriminate between cyberattack events and natural power system disturbances which yielded improved attack detection performance and improved robustness versus stragglers.

Whilst conventional FL has proved its effectiveness in the area of anomaly detection for power systems, it is accompanied with inherent limitations that necessitate further investigations. One such major drawback is its heavy reliance on a centralized aggregator which introduces potential privacy and security hazards during the transmission of sensitive model updates to and from client nodes. Furthermore, the communication overhead associated with the bi-directional transmission of model gradients from numerous edge devices to the central aggregator can be computationally expensive, thus leading to scalability issues, particularly in scenarios with a large number of client nodes or when operating in resource-constrained environments. Hence, it is imperative to devise a solution to address these technical bottlenecks.

\subsection{Decentralized Federated Anomaly Detection}

In the context of power system scenarios, the novelty of Decentralized Federated Learning (DFL) emerges as a vital solution to overcome the aforementioned technical challenges associated with conventional FL-based anomaly detection. DFL is a recently surfaced machine learning paradigm which combines the principles of FL with a decentralized architecture. In conventional FL, a central aggregator orchestrates the model training process by aggregating model updates from multiple client nodes. In contrast, model training is distributed across multiple client nodes during DFL. Each node independently trains a local model using its own data and periodically exchanges model updates or gradients with other client nodes within the network. In the context of power systems, Giuseppi et al. \cite{9837291} devised a decentralized version of FederatedAveraging (DecFedAvg) algorithm for non-intrusive load monitoring in smart energy communities and validated the effectiveness of their innovative approach using REFIT dataset which yielded low Mean Absolute Error (MAE) values. Another work by Gao et al. \cite{10.1145/3485730.3493450} presented a DFL-based residential building load forecasting framework with a gradient selection mechanism and achieved a 97\% accuracy on Pecan Street Dataset. Remarkably, only a limited number of studies have explored the use of DFL within the smart grid domain with a notable absence of studies specifically addressing the issue of anomaly detection. This evident gap in existing literature highlights the need for additional investigation, prompting our motivation. Therefore, this manuscript aims to address this gap by proposing a novel approach to anomaly detection in power systems by decentralizing the FL process using a gossip-based approach to eliminate the dependency on a centralized aggregator.

\section{Proposed Method}
\label{sec:proposedmethod}
Throughout this section, we will initially present the system model considered for this scenario followed by a discussion of our proposed DFL-based anomaly detection framework.

\subsection{Power System Model \& Assumptions}

\begin{figure}
    \centering
    \includegraphics[width=9cm]{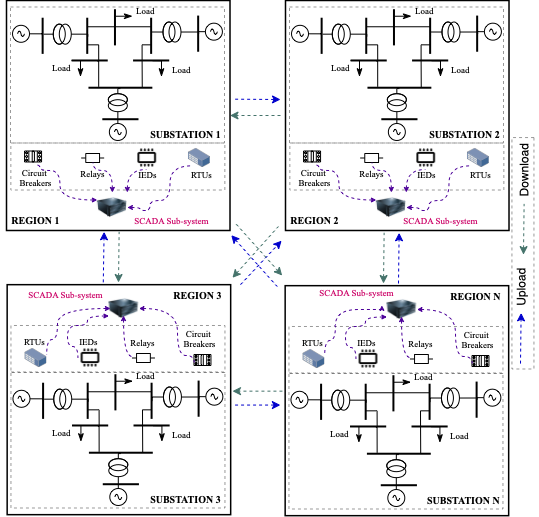}
    \caption{A visual representation of our proposed DFL-based anomaly detection power system model. An intuitive workflow of our proposed framework is as follows: 1) Model initialization on each participating node. 2) Local training of models on each node. 3) After training, each node sends its local model updates to peers throughout the network. 4) After receiving model updates from its peers, each node aggregates these updates with its own local model parameters. 5) The steps 2-4 are repeated until model maturity is achieved.}
    \label{fig:systemmodel}
\end{figure}

Throughout this manuscript, we consider a geographically decentralized power grid system that is segmented into $N$ number of grid regions where $N \in \mathbb{R}^+$. As depicted in Fig \ref{fig:systemmodel}, we briefly introduce the primary actors of power system model as follows: 1) \textit{SCADA Sub-systems}: The SCADA sub-systems are the edge-based client nodes equipped with sophisticated data acquisition technologies which play a crucial role in gathering and overseeing the data sensed by each sensing device (including relays, Remote Terminal Units (RTUs), etc.) within a power grid region. Furthermore, the SCADA sub-systems will function as DFL nodes whereby they will each train a model locally using their own data repository and update the compressed gradients to selected peers within the network. Lastly, the SCADA sub-systems are responsible for the aggregation of the received updates from their peers with their own local model parameters. For this given scenario, we assume that the SCADA sub-systems are honest-but-curious, i.e, they strictly adhere to the protocols of DFL without exhibiting malicious behaviours to ensure data integrity and privacy, model quality, fairness, etc.  2) \textit{Grid Regions}: The power grid regions are equipped with a variety of decentralized monitoring sensoring devices including relays, synchrophasors, RTUs, etc. and are responsible for the continual sensing of grid-related data such as current phase angle, voltage magnitude, etc. The Intelligent Electronic Devices (IEDs) are presumably connected to the SCADA sub-systems via high speed data transmission networks built upon DNP3/IEC 61850 protocols. Furthermore, we assume that the grid regions are interconnected for operational purposes using wireless mesh networking protocols including Zigbee, etc.

\subsection{Proposed Anomaly Detection Mechanism}

\begin{figure*}
    \centering
    \includegraphics[height=4.7cm, width=16cm]{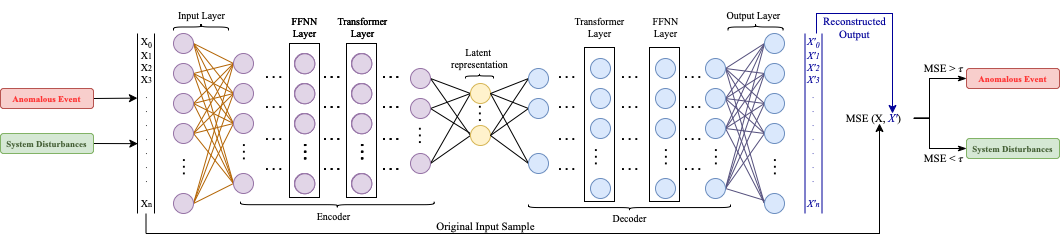}
    \caption{An overview of our anomaly detection module accompanied by an illustration of an input vector undergoing the anomaly detection process. It consists of three elements: 1) encoding phase, 2) latent representation phase and 3) decoding phase.} \label{fig:taemodel}
\end{figure*}

As previously mentioned in Section \ref{sec:relatedliterature} above, previous works in similar field have employed a number of anomaly detection techniques including statistical approaches, traditional ML algorithms, DL neural networks, etc. However, in this manuscript, we formulate the problem of cyber attack detection using a Transformer-Autoencoder (TAE) model as illustrated in Fig. \ref{fig:taemodel}. TAEs are powerful hybrid neural networks that combines the elements of both transformer and autoencoder models, leveraging the strengths of sequence modelling and representation learning, with the objective of achieving identity mapping between inputs and outputs, whilst capturing complex patterns and dependencies from the data \cite{aspect_akbar}. Autoencoders learn representations in data by passing an input data vector $v_{i}$ through an encoder $f_\alpha$ which maps each input $v_{i}$ to a latent space representation $r^{(j)}$ which is essentially a low-dimensional and compressed knowledge representation of the original input such that $r^{(j)} = f_\alpha (v_i) = a(Wv_i + b)$ where $a$ is an element-wise activation function, $W$ is the weight matrix and $b$ is the bias vector of the encoder. On the other hand, the decoder $g_{\alpha}$ is responsible for the reconstruction of the original input $v_i$, with minimal reconstruction error, from $r^{(j)}$ through $v_j = g_\alpha (v_i) = a(W'v_i + b')$ where $v_j$ is the reconstructed input (output of decoder). During the federated training process, the model parameters $W$, $W'$, $b$ and $b'$ are iteratively improved to optimize the reconstruction error $\ell (v_i, v_j)$ such that $\alpha , \alpha' = argmin_{\alpha, \alpha'} \ell (v_i, v_j)$.

Our proposed TAE model architecture consists of two symmetrical components, namely, the encoder and the decoder, each comprising of a mixture of Feed Forward Neural Network (FFNN) layers and transformer layers, which utilize self-attention mechanisms to capture dependencies between input tokens and model long-range interactions. These transformer layers enable effective sequence modeling and feature extraction which is crucial for identifying complex patterns indicative of anomalies in the data. Each layer within the TAE facilitates communication between its preceding and succeeding layers, serving as a hidden layer to the previous nodes and a visible one to the subsequent nodes. During learning, the parameters of the model, including weight and bias, are adjusted to minimize the reconstruction error, which serves as an indication of the existence of anomalies. Leaky Rectified Linear Unit (ReLU) is used as the activation function for its robustness against noisy and sparse data while Mean Squared Error (MSE) is chosen as the reconstruction error metric for the measurement of anomaly degree. An anomaly threshold $\tau$ is established by initially sorting the anomaly degrees in increasing order and selecting the optimal value of $\tau$ at the point of inflection in the error distribution. Furthermore, soft-max layers are incorporated at the end of the TAE layers stack to enable the prediction of attack labels based on learned representation. The training batch size and the learning rate are both configured to $100$ and $10^{-3}$ respectively. This model serves as the cornerstone for power systems anomaly detection in this manuscript.

\subsection{Gradient Quantization}

During federated training, the canonical objective of the TAE model is minimize the objective function $argmin_{\alpha, \alpha'} \ell (v_i, v_j)$. Typically, the search for the optimal parameter values of this objective function within a FL/DFL set-up can be accomplished by employing several optimization algorithms including Stochastic Gradient Descent (SGD), Adam, etc. However, due to the requirement of  bi-directional communication amongst grid regions, the size and frequency of the local gradient updates increases which hinders the scalability of our proposed approach to increasing number of grid regions. Additionally, the sharing of raw model updates poses significant privacy risks to such a critical system, this non-conformant to confidentiality guarantees of FL \cite{10173657}. 

Therefore, to overcome such technical bottlenecks, we propose to employ a gradient quantization scheme, specifically DP-SIGNSGD \cite{feddisc}, which approximates the TAE model gradients into low-precision values to reduce communication bandwidth. After aggregation with previous model updates, each decentralized grid region $k$ executes a zero-mean normalization of its computed gradient $g_k^t$ such that $\overline{g}_k^t = g_k^t - \mu_k^t$ for uniform distribution, followed by a compression of the normalized gradients by each grid region $k$ through a one-bit differentially-private quantizer which eventually captures the sign of $\overline{g}_k^t$, such that $\hat{g}_k^t = dpsign(\overline{g}_k^t)$. The quantized gradient $\hat{g}_k^t$ from one grid region is then sent to a randomly selected neighbour node. The latest model gradient is achieved by updating the new weight parameters such that $W^{t+1} = W^{t} - \eta^t\hat{g}$ where $\eta^{t} \in (0,1)$ represents the learning rate. This gradient quantization technique, as detailed in Algorithm \ref{algo:dfl}, significantly decreases the byte size of the gradient updates exchanged during the gossip-based model training. 

\subsection{Gossip-based DFL Framework}

As presented in Fig. \ref{fig:systemmodel}, throughout our manuscript, we focus a geographically decentralized region-wise power grid system which is interconnected with wireless technologies. The primary objective of the decentralized federated anomaly detection scenario is to collaboratively train the global TAE model across the participating grid clients without the need to transmit local model updates to a centralized server for aggregation. Let $D_k = \{a_k^i, b_k^i\}^{N_k}_{i=1}$ denote the training dataset for a specific grid region $k \in [K]$ where $N_k = |D_k|$ represents the number of data points for each $D_k$. It is crucial to note that the data distributions are Non-IID, mirroring real-world intrusion detection scenarios. Hence, the objective function $f_k$ for a shared global TAE model parameter $W \in \mathbb{R}^M$ can be expressed as $f_k(W) = \dfrac{1}{N_k} \displaystyle \sum_{i=1}^{N_k}\ell \left(\{a_k^i, b_k^i\}; Z \right)$. Here, $\ell (.): \mathbb{R}^M \times \mathbb{R} \rightarrow \mathbb{R}$ represents the loss function to be optimized for data point $\left(a_k^i, b_k^i\right)$. In addition, the global objective function $F$ across all $k$ can be expressed as $F(W) = \displaystyle \sum_{k=1}^K \dfrac{N_k}{N}f_k(W)$ which is the sum of the local loss functions for each $k$, where $N$ is the total data points for training.

To enable federated discrimination of attack events from non-anomalous ones, we propose to implement two types of gossip protocols: 1) \textit{Random Gossip} randomly selects client nodes within a network for communication. During each communication round, each $k$ computes its local gradient $\nabla f_k(Z^t)$ using its training dataset $D_k$ and aggregates this gradient with that received from a randomly chosen peer node. This aggregated gradient is then normalized and quantized before being transmitted to another randomly selected peer node. This process repeats iteratively, facilitating decentralized communication and collaboration among nodes. 2) \textit{Epidemic Gossip}, on the other hand, operates on the principle of spreading information to all client nodes within a network. Each $k$ communicates with a fixed subset of its neighboring $k$, known as its Epidemic Radius. Similar to random gossip, nodes compute, aggregate, normalize, and quantize gradients before transmitting them to neighboring nodes. However, unlike random gossip, Epidemic gossip ensures that each node eventually receives the information from all other nodes in the network. By leveraging random gossip, nodes can efficiently exchange information without requiring explicit knowledge of the network topology. This enhances scalability and adaptability to changes in network conditions. On the other hand, Epidemic gossip ensures that all nodes eventually receive the information, thereby promoting information dissemination and consensus among nodes. 

During a particular communication round $t^{th}$, our proposed decentralized federated anomaly detection, as presented in Algorithm \ref{algo:dfl} constitutes of five specific implementation steps as discussed in the following: 1) Each grid region $k \in [K]$ computes its local gradient $g^t_k = \nabla f_k(W^t)$ using its training dataset $D_k$ and the current global model parameter $W^t$. 2) Each $k$ then performs a majority-vote based aggregation and summation of the computed gradient $g^t_k$ along with the previous gradient $g^t_{k-1}$ received from the previous peer such that $\displaystyle \sum_{i=k-1}^{k}g^t_{agg}$. 3) The aggregated gradient $g^t_{agg}$ is then normalized through zero-mean normalization such that $\bar{g}^t_{norm} = g^t_{agg} - \mu^t_k$ to ensure consistency, stability and improved convergence. 4) The normalized gradient $\bar{g}^t_{norm}$ is then quantized using the differentially private quantizer $dpsign(\cdot)$ to obtain the quantized gradient $\hat{g}^t_k$. 5) Lastly, the quantized gradient $\hat{g}^t_k$ is then updated to a peer node within the network. In case of Random Walk gossip, each $k$ transmits $\hat{g}^t_k$ to another randomly selected peer node and during Epidemic gossip, each $k$ communicates with a fixed subset of its neighboring $k$.


\begin{algorithm}
\textbf{Input}: $K$ Grid Regions, one-bit quantizer $dpsign(.)$, Local training set $D_k$ for $k \in [1, K]$, learning rate $\eta$,
\vskip0.5em
\hskip0.5em \textit{Initialize} global unanimous model $W_0$.
\bigbreak

\hskip0.5em \For{each communication round $t$ in $T_{cl} \in (1,n)$}{

\hskip0.5em \For{each $k \in [K]$}{

\textit{Calculate} local gradient $g^t_k = \nabla f_k(W^t)$ using $D_k$. 
\vskip0.5em
\textit{Aggregate} local gradient $g^t_k$ with previous peer, $g^t_{k-1}$ such that $\displaystyle \sum_{i=k-1}^{k}g^t_{agg}$..
\vskip0.5em
\textit{Calculate} normalized gradient $\bar{g}^t_{norm} = g^t_{agg} - \mu^t_k$.
\vskip0.5em
\textit{Calculate} quantized gradient $\hat{g}^t_k = dpsign$ $(\bar{g}^t_{norm})$.
\vskip0.5em
\textit{Communicate} $\hat{g}^t_k$ to peer node.
\vskip0.5em
\textbf{end}
}

}
\textbf{end}
\caption{Proposed Gossip-based DFL Framework}
\label{algo:dfl}
\end{algorithm}


\section{Empirical Evaluation \& Discussions}
\label{sec:results}
In this section, we initially discuss the dataset used for the experimental set-up followed by comprehensive discussion of the empirical results which validate the anomaly detection performance and robustness of our proposed solution as presented in Section \ref{sec:proposedmethod}.

\subsection{Dataset Description \& Preparation}
We conduct an empirical evaluation of our proposed federated decentralized anomaly detection framework for power systems using the openly available Mississippi State University and Oak Ridge National Laboratory Power System Attack (MSU-ORNL PSA) dataset, which is derived from industrial control systems [16]. This dataset consists of 15 distinct data files, each representing various power system event scenarios. These scenarios encompass both natural events, such as power system disturbances, and deliberate attack events, including cyberattacks. Natural events are categorized based on fault severity across distributed grids, ranging from 10-19\%, 20-79\%, to 80-90\%. Conversely, attack events include: 1) remote tripping command injection, involving sending commands to IEDs to manipulate circuit breakers; 2) changes in IED settings to disrupt normal operations; and 3) false data injection attacks, intelligently altering power system values across parameters like voltage and current to mislead system operators. The datasets consist of 128 features, encompassing measurements e.g. frequency, voltage phase angle and current magnitude, amongst others.

We begin by selecting a basis of 100 necessary features for training our TAE-based anomaly detection neural network, a selection made following Principal Component Analysis (PCA) to cull out redundant features and decrease training complexity. To address the issue of missing values within the dataset, we employ K-Nearest Neighbor (KNN) imputation to prevent information loss as opposed to downsampling. Furthermore, we normalize the input features to fall within the range of (0,1). Finally, we partition the dataset into training and testing sets using a split ratio of 80\% for training and 20\% for testing, respectively.

\subsection{Anomaly Detection Performance}

\begin{table}
\centering
\begin{tabular}{|l|c|c|c|c|c|}
\hline
\diagbox{\textbf{Performance Metric}}{\textbf{Learning Model}} & \textbf{Proposed Approach} & \textbf{CNN} & \textbf{LSTM} & \textbf{RNN} & \textbf{RBM} \\ \hline
\textbf{Anomaly Detection Accuracy} & 0.942 & 0.915 & 0.819 & 0.751 & 0.798 \\ \hline
\textbf{Anomaly Detection Precision} & 0.926 & 0.909 & 0.784 & 0.728 & 0.784 \\ \hline
\textbf{Anomaly Detection Recall} & 0.929 & 0.907 & 0.778 & 0.723 & 0.779 \\ \hline
\textbf{Anomaly Detection F-Score} & 0.937 & 0.912 & 0.791 & 0.725 & 0.782 \\ \hline
\end{tabular}
\caption{Comparison of average anomaly detection performance of our proposed TAE model against other neural networks in a Random Walk Gossip scenario. Our proposed TAE model consistently outperforms all other models in terms of anomaly detection performance.}
\label{table:metricperformcompaRW}
\end{table}

\begin{table}
\centering
\begin{tabular}{|l|c|c|c|c|c|}
\hline
\diagbox{\textbf{Performance Metric}}{\textbf{Learning Model}} & \textbf{Proposed Approach} & \textbf{CNN} & \textbf{LSTM} & \textbf{RNN} & \textbf{RBM} \\ \hline
\textbf{Anomaly Detection Accuracy} & 0.915 & 0.914 & 0.820 & 0.752 & 0.799 \\ \hline
\textbf{Anomaly DetectionPrecision} & 0.897 & 0.910 & 0.785 & 0.729 & 0.785 \\ \hline
\textbf{Anomaly DetectionRecall} & 0.898 & 0.906 & 0.779 & 0.724 & 0.780 \\ \hline
\textbf{Anomaly DetectionF-Score} & 0.906 & 0.913 & 0.792 & 0.726 & 0.783 \\ \hline
\end{tabular}
\caption{Comparison of average anomaly detection performance of our proposed TAE model against other neural networks in an Epidemic scenario. Our proposed TAE model consistently outperforms all other models in terms of anomaly detection performance.}
\label{table:metricperformcompaEpi}
\end{table}

\begin{figure}[h]
    \centering
        \centering
        \includegraphics[width=7.5cm]{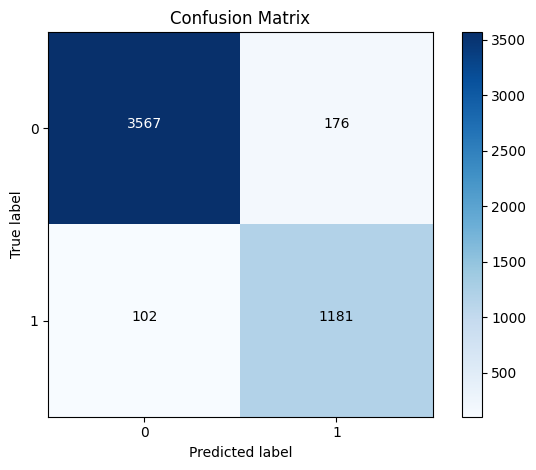}
        \label{fig:EpiConfu}
        \centering
        \includegraphics[width=7.5cm]{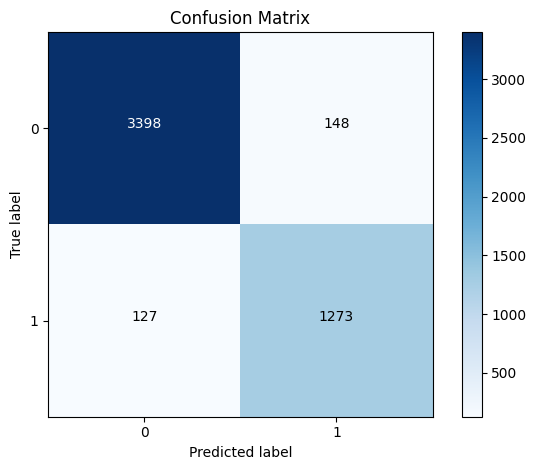} 
        \label{fig:RWconfu}
    \caption{Comparison of confusion matrices for Epidemic (above) and RandomWalk (below) pertaining to a single data file evaluated using our proposed DFL approach. Upon analyzing the occurrences of false positives and false negatives, it becomes evident that Random Walk gossip DFL outperforms Epidemic gossip DFL approach.}
    \label{fig:confusion matrix}
\end{figure}

\begin{figure}
    \centering
    \includegraphics[width=7.5cm]{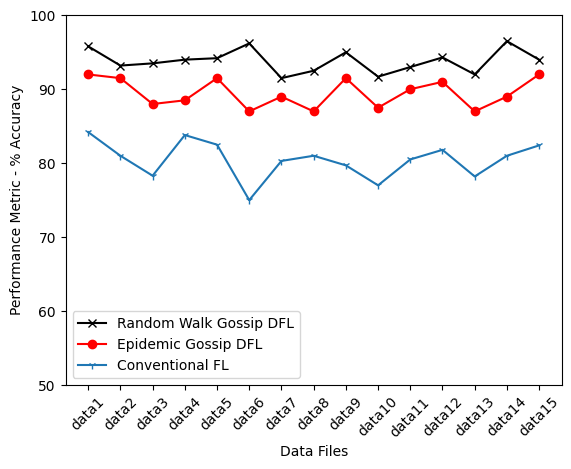}
    \caption{Comparative Analysis of the performance of our two proposed gossip-based DFL approaches against a conventional FL solution over the 15 data files of the MSU-ORNL dataset. Notably, the results indicate the superiority of Random Walk Gossip-based DFL solution against other setups. }
    \label{fig:datafil}
\end{figure}

Table \ref{table:metricperformcompaRW} and Table \ref{table:metricperformcompaEpi} demonstrate the performance metrics of the two gossip-based decentralized federated learning approaches applied to anomaly detection in power systems, with Table \ref{table:metricperformcompaRW} representing the results obtained using the random walk gossip-based method Table \ref{table:metricperformcompaEpi} corresponding to the epidemic gossip-based approach. It is clear from comparing the two tables' results that the random walk gossip-based approach outperforms the epidemic gossip-based method consistently in terms of all performance criteria. The random walk gossip-based approach in DFL encourages efficient communication and information exchange among grid regions by allowing each client node to randomly select its peer. This randomness encourages a diverse exchange of model updates,thus enhancing convergence and accuracy of model updates. Furthermore, a balanced distribution of updates is achieved, thus mitigating information bottlenecks and ensuring all grid regions contribute effectively to the learning process. Similarly, based on Fig. \ref{fig:confusion matrix} whereby the confusion matrices depict the performance of the Random Walk gossip-based (above) and Epidemic gossip-based (below) approaches, it is evident based on analysis of the true positive and true negative values that Random Walk gossip-based DFL approach exhibits superior performance. The higher true positive and true negative rates in the Random Walk gossip-based DFL matrix suggest that it more accurately identifies and correctly classifies both positive and negative instances, making it a preferable choice for our particular scenario. Furthermore, in Fig. \ref{fig:datafil}, we contrast the anomaly detection performance of the two gossip based protocols against that of conventional FL. The conventional FL approach demonstrates the lowest performance due to its inherent limitations in facilitating efficient communication and information exchange among the distributed nodes. 

\subsection{Computational Overhead Analysis}

\begin{figure}
    \centering
    \includegraphics[width=7.5cm]{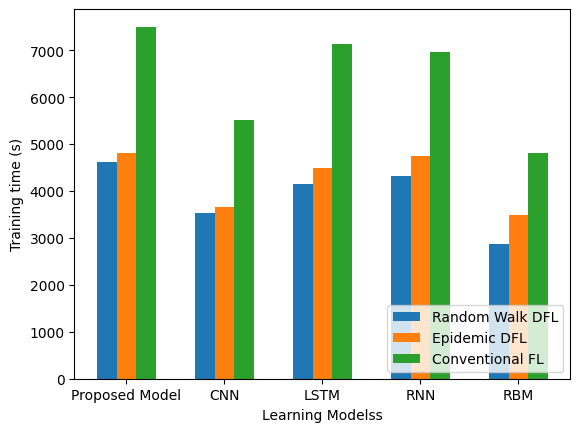}
    \caption{Comparison of time taken to train our proposed DFL-based anomaly detection solution while employing two distinct gossip-based protocols against a conventional Federated Learning setup.}
    \label{fig:overheadanalysis}
\end{figure}

Following the successful validations of the anomaly detection performance of our proposed approach, we further assess the the computational overhead achieved through elimination of the centralized aggregator and gradient quantization as presented in Section \ref{sec:proposedmethod}. Specifically, we contrast the time taken to train the neural networks in three distinct scenarios namely Random Walk gossip-based DFL, Epidemic gossip-based DFL and conventional FL. To ensure fairness during comparison, we utilize similar models with similar hyperparameters in all three setups. Fig. \ref{fig:overheadanalysis} illustrated a comparison of the computational efficiency among the two aforementioned types of gossip DFL protocols and a conventional FL scenario. The results reveal a consistent decrease in training time across all models when our proposed DFL-gradient quantization approach is employed against a conventional FL scheme. Notably, the Random Walk and Epidemic Gossip based FL approach demonstrate approximately 35\% less training time compared to the traditional FL framework. Furthermore, it is evident that Random Walk-based DFL demonstrates around 3\% to 5\% reduction in training time as compared to Epidemic-based DFL, indicating that Random Walk is less computationally demanding and much better suited for resource-constrained environments such as in the case of power systems. 

\subsection{Summary of findings}
In this section, we investigated the detection performance of our novel decentralized anomaly detection solution within power systems, focusing on the comparison between decentralized federated learning (DFL) and conventional federated learning (FL) methodologies. Through extensive experimentation and analysis, we found that DFL methods by employing gossip-based communication protocols such as random walk and epidemic gossip, consistently outperformed the conventional FL scheme. Specifically, DFL methods exhibited superior detection accuracy and more efficient communication and information exchange among participating grid regions. The decentralized nature of DFL enables more dynamic and adaptive learning processes, leveraging the collective intelligence of distributed grid regions while mitigating the limitations associated with centralized coordination mechanisms in conventional FL setups. Additionally, we investigated the computational overhead and found that the conventional FL method incurred higher time costs compared to DFL approaches, further emphasizing the computational efficiency and scalability of decentralized federated anomaly detection for SGs.

\section{Conclusion}
\label{sec:conclusion}

Throughout this manuscript, we investigated the effectiveness of a Decentralized Federated Learning (DFL) scheme for anomaly detection in the domain of power systems. To the best of our knowledge, we are the first to introduce a gossip-based solution to decentralized federated anomaly detection in the SG ecosystem. First, we proposed two gossip protocols namely Random Walk and Epidemic that enables the transition from a conventional Federated Learning paradigm to a decentralized one while eliminating the need for a centralized aggregator. Secondly, we proposed a hybrid anomaly detection neural network namely a Transformer Autoencoder (TAE) model. Extensive experimental validations using a publicly available dataset show that the TAE achieves superior accuracy as opposed to other state-of-the-art benchmark neural networks. Lastly, we introduce a sign-based gradient quantization scheme to optimize communication efficiency within the DFL framework, resulting in a significant reduction in training time. Overall, the findings concluded that there is a compelling potential for developing anomaly detection solutions using DFl and that Random Walk has shone in terms of both detection performance and computational overhead.

In future works, our main objective would be to improve the resilience of intrusion detection models by exploring more complex hybrid techniques within the DFL paradigm. Additionally, we will be investigating on the different cyberattacks e.g. Byzantine attacks that may disrupt the operations of the DFL framework.

\bibliographystyle{cas-model2-names}

\bibliography{cas-refs}

\end{document}